\documentclass[%
 reprint,
superscriptaddress,
 amsmath,amssymb,
 aps,
]{revtex4-1}

\usepackage[pdftex]{graphicx}
\usepackage{dcolumn}
\usepackage{bm}
\usepackage{textcomp}

\begin{document}


\title{Ultrahigh-\textit{Q} crystalline microresonator fabrication with only precision machining}

\author{Shun Fujii}
\affiliation{Department of Electronics and Electrical Engineering, Faculty of Science and Technology, Keio University, Yokohama, 223-8522, Japan}

\author{Yuka Hayama}
\affiliation{Department of System Design Engineering, Faculty of Science and Technology, Keio University, Yokohama, 223-8522, Japan}

\author{Kosuke Imamura}
\affiliation{Department of System Design Engineering, Faculty of Science and Technology, Keio University, Yokohama, 223-8522, Japan}

\author{Hajime Kumazaki}
\affiliation{Department of Electronics and Electrical Engineering, Faculty of Science and Technology, Keio University, Yokohama, 223-8522, Japan}

\author{Yasuhiro Kakinuma}
\affiliation{Department of System Design Engineering, Faculty of Science and Technology, Keio University, Yokohama, 223-8522, Japan}

\author{Takasumi Tanabe}
\email{takasumi@elec.keio.ac.jp}
\affiliation{Department of Electronics and Electrical Engineering, Faculty of Science and Technology, Keio University, Yokohama, 223-8522, Japan}


%

\date{\today}

\begin{abstract}
The development of ultrahigh quality factor (\textit{Q}) microresonators has been driving such technologies as cavity quantum electrodynamics (QED), high-precision sensing, optomechanics, and optical frequency comb generation. Here we report ultrahigh-\textit{Q} crystalline microresonator fabrication with a \textit{Q} exceeding $10^8$, for the first time, achieved solely by computer-controlled ultraprecision machining. Our fabrication method readily achieved the dispersion engineering and size control of fabricated devices via programmed machine motion. Moreover, in contrast to the conventional polishing method, our machining fabrication approach avoids the need for subsequent careful polishing, which is generally required to ensure that surface integrity is maintained, and this enabled us to realize an ultrahigh-\textit{Q}. We carefully addressed the cutting condition and crystal anisotropy to overcome the large surface roughness that has thus far been the primary cause of the low-$Q$ in the machining process. Our result paves the way for future mass-production with a view to various photonic applications utilizing ultrahigh-\textit{Q} crystalline microresonators.
\end{abstract}

\maketitle


\section{Introduction}

Ultrahigh-$Q$ crystalline microresonators have been used as attractive platforms for studying nonlinear and quantum optics in the last few decades~\cite{PhysRevA.74.063806,GRUDININ200633,Lin:17,doi:10.1002/lpor.201600038,Kovach:s}. In particular, laser stabilization via self-injection locking and Kerr optical frequency comb generation are potential applications with the aim of realizing an optical-frequency synthesizer~\cite{Lucas2020} and low-noise, compact photonic devices~\cite{Liang2015,Pavlov2018}. Injection locking to high-$Q$ whispering gallery mode (WGM) microresonators enables the laser linewidth to be reduced to less than hundreds of hertz~\cite{Liang:10}. Moreover, Kerr frequency comb generation~\cite{del2007optical} provides RF oscillators with high spectral purity~\cite{Liang2015:high,PhysRevLett.122.013902} and an optical pulse train with a high repetition rate~\cite{herr2014temporal,yi2015soliton}. These applications rely on the high-$Q$ of crystalline microresonators, typically up to $10^9$ and corresponding to a resonance linewidth of hundreds of kilohertz, which enhances the optical nonlinearity. The fundamental limit of the $Q$-factor in crystalline resonators is $\sim$$10^{13}$~\cite{Grudinin:07} ($Q>10^{11}$ as observed in the experiment~\cite{Savchenkov:07}), and this value surpasses that of resonators made with other materials (e.g., silica, silicon, etc)~\cite{PhysRevA.70.051804}. In addition, they have a fully transparent window in the visible to mid-infrared wavelength region, which expands the available bandwidth as well as the telecom band~\cite{Lin:15}.

Magnesium fluoride ($\mathrm{MgF_2}$) and calcium fluoride ($\mathrm{CaF_2}$) are crystalline materials that are commonly used for fabricating WGM microresonators thanks to their quality, commercial availability, and optical properties. We usually manufacture crystalline resonators by using diamond turning and a polishing process.  They are accomplished either with a motion-controlled machine or manually~\cite{Savchenkov:06}. A hard diamond tool enables us to fabricate WGM structures, but we have to employ subsequent manual polishing with diamond slurry to improve the $Q$-factor of the microresonator. Precision machining readily overcomes the geometrical limitation of the manual process; therefore, precise computer-controlled machining has achieved the pre-designed mode structures needed when fabricating single-mode~\cite{Savchenkov:06,Winkler:16} and dispersion engineered resonators to generate broadband microresonator frequency combs ~\cite{Grudinin:12,Grudinin:15,Nakagawa:16}. However, a significant challenge remains because we need to employ additional hand polishing after the diamond turning process due to the relatively low $Q$ of $10^6$$\sim$$10^7$ at best that we obtain when using machining alone~\cite{Savchenkov:06,Grudinin:12,Grudinin:15,Pavlov:17}. The additional polishing improves the $Q$; however, subsequent polishing deforms the precisely fabricated structures despite the engineered dispersion realized by the programmed motion of the lathe~\cite{Savchenkov:06}.

In this article, we describe an ultrahigh-$Q$ crystalline microresonator fabrication technique that employs computer-controlled ultraprecision machining. The measured $Q$ of the $\mathrm{MgF_2}$ crystalline resonator reaches $1.4\times10^8$, which is the highest value yet obtained without a subsequent polishing process. In addition, we achieved a comparably high-$Q$ in $\mathrm{CaF_2}$ crystalline material. To obtain the ultrahigh-$Q$, we addressed the single-crystal cutting condition by undertaking an orthogonal cutting experiment, which revealed the critical depths of cuts for different cutting directions. Also, a precise cylindrical turning experiment revealed the relationship between crystal anisotropy and surface quality after machining and demonstrated the realization of nanometer-scale surface roughness with diamond turning alone. The results we obtained provide clear evidence that cutting parameters that have been optimized for fluoride crystals lead to a significant reduction in surface roughness.

An automated ultra-precision machining technique is compatible with dispersion engineering and a high-$Q$ factor, which is often restricted by the trade-off relation encountered with conventional fabrication techniques. We confirmed that the dispersion of fabricated resonators agrees extremely well with the design, and this extends the potential of dispersion controllability in crystalline WGM microresonators. Furthermore, our approach, namely the reliable production of high-$Q$ crystalline microresonators, supports recent advances on the integration of crystalline microresonators with photonic waveguides towards a wide range of future applications~\cite{Anderson:18,Liu:18}.

\section{Investigation of critical depth of cut}

\begin{figure}[p]
	\centering
	\includegraphics[]{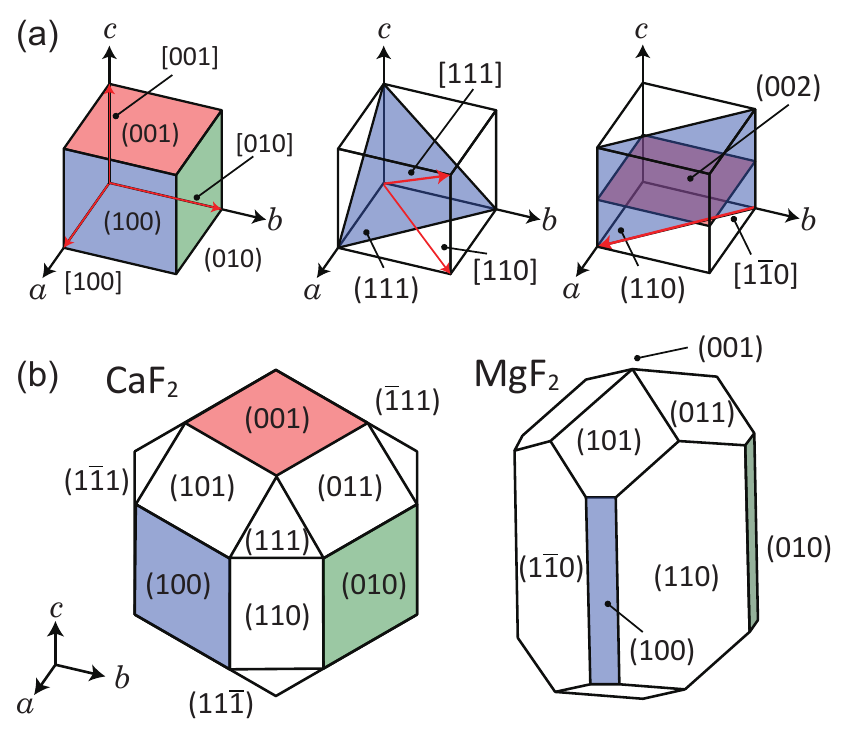}
	\caption{(a) Examples of the Miller index, where ($h~k~l$) and [$h~k~l$] indicate the corresponding plane and direction, respectively. (b) Crystallographic images of $\mathrm{CaF_2}$ and $\mathrm{MgF_2}$ crystal. In contrast to the cubic symmetry system of $\mathrm{CaF_2}$, $\mathrm{MgF_2}$ is characterized by a more complex rutile structure.
	}
	\label{fig1}
\end{figure}

\begin{figure*}[b]
	\centering
	\includegraphics[]{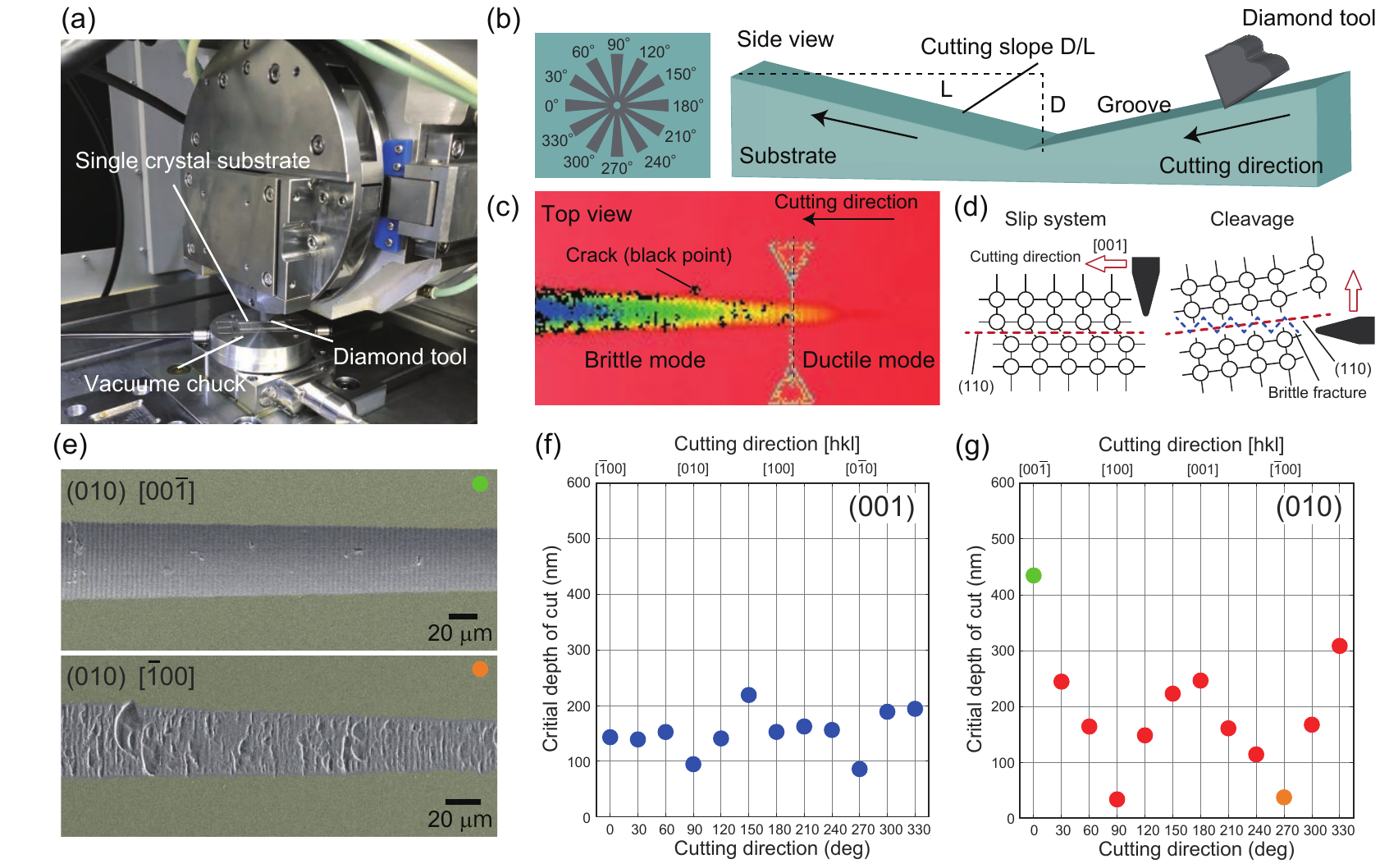}
	\caption{(a) Experimental setup for orthogonal cutting to investigate the critical depth of cut in $\mathrm{MgF_2}$ single crystal. (b) Schematic illustration of the diamond tool to substrate motion. The diamond tool cut a V-shaped groove with a slope of $D/L$. (c) Reconstructed image of machined surface using scanning white light interferometer. The critical depth of cut is given by the depth where the first brittle fracture appeared (black point in (c)). (d) Cutting along the slip plane (110)[001] promotes ductile mode cutting (left panel). Cleavage and subsequent brittle fractures are induced by the cutting force against cleavage plane (right panel). (e) Scanning electron micrographs showing machined surfaces of a (010) plane with [00$\overline{1}$] direction (upper panel) and [$\overline{1}$00] direction (lower panel). The yellow shaded area corresponds to the original uncut surfaces. The difference between the machined surface conditions is attributed to the crystal anisotropy of the $\mathrm{MgF_2}$ crystal. (f), (g) The measured critical depth of cut versus cutting direction on a (001) and (010) plane, respectively. In comparison with the (001) plane, the (010) plane shows large variation in cutting direction due to crystal anisotropy. To perform ductile mode cutting, the depth of cut must be kept below the critical depth of cut.}
	\label{fig2}
\end{figure*}

\subsection{Definition of critical depth of cut}

One important parameter that we need to know when fabricating a single crystal is the critical depth of cut. It is defined by the depth of cut at which the transition from ductile-mode to brittle-mode cutting is observed when machining single-crystal material~\cite{doi:10.1111/j.1151-2916.1990.tb05142.x}. In the ductile regime, a smooth crack-free surface can be maintained when generating a continuous ribbon chip, and this approach is considered more suitable for optical applications thanks to its ultra-smooth surface. On the other hand, the surface in the brittle regime is rougher and contains cracks, hence it is generally inadequate for optical applications. $\mathrm{CaF_2}$ and $\mathrm{MgF_2}$ crystals are hard and brittle materials, and have a crystal anisotropy, so they are challenging to cut. These features make it difficult to manufacture smooth optical elements with a designed shape such as spherical lenses and optical microresonators. In particular, high-$Q$ microresonators require an ultra-smooth surface with a surface roughness of no more than a few nanometers. Thus, the critical depth of cut must be investigated before resonator fabrication if we are to cut the crystal in the ductile mode regime.

The cutting direction and plane of the crystal is expressed with the Miller index, as shown in Fig.~\ref{fig1}(a) (Details of the Miller index and crystallographic structure can be found in Supplement~1). Figure~\ref{fig1}(b) shows crystallographic images of $\mathrm{CaF_2}$ and $\mathrm{MgF_2}$ crystal. The difference in crystal structure influences the critical depth of cut and the cutting conditions.

\subsection{Orthogonal cutting experiment}

To investigate the critical depth of cut, we performed an orthogonal cutting experiment on single-crystal $\mathrm{MgF_2}$. The experiment was carried out with an ultra-precision machining center (UVC-450C, TOSHIBA MACHINE), and a workpiece holder equipped with a dynamometer to detect the cutting force during the processing. As a workpiece, we used a pre-polished single-crystal $\mathrm{MgF_2}$ substrate with a size of 38$\times$13~mm and a thickness of 1~mm, which was fixed on the workpiece holder with a vacuum chuck as shown in Fig.~\ref{fig2}(a). The cutting tool was a single crystal diamond tool with a 0.2~mm nose radius, a $-20^\circ$ rake angle, and a 10$^\circ$ clearance angle (Details of the single-crystal diamond tool are provided in Supplement~1). The cutting slope $D/L$, which gives the cutting depth to cutting length ratio, and the feed rate, were set at 1/500 and 20~mm/min, respectively, with a numerical control (NC) program [Fig.~\ref{fig2}(b)]. The critical depth of cut, which is defined as the cutting depth at which the first brittle fracture appeared on the surface, was measured using a scanning white light interferometer (New View TM6200, Zygo). Figure~\ref{fig2}(c) shows an image of the machined surface, where the black points indicate fractures or cracks on the surface.

Here, we tested two different crystal planes, (001) and (010), where we performed the cutting in every 30$^\circ$ rotational direction to investigate the critical depth of cut for different crystal orientations. The direction of 0$^\circ$ was set at [$\overline{1}$00] and [00$\overline{1}$], respectively. It should be noted that $\mathrm{MgF_2}$ has a complex rutile structure with a different crystal plane configuration from $\mathrm{CaF_2}$ [see Fig.~\ref{fig1}(b)]; hence the two orthogonal planes are selected for the test to reveal the effect of crystal anisotropy. Figure~\ref{fig2}(d) shows a schematic of the tensile stress model for single-crystal cutting, and this will be explained in more detail later.

Figure~\ref{fig2}(e) presents scanning electron microscope (SEM) images showing the surface condition of the (010) plane after orthogonal machining (the yellow region is the original uncut surface). Even though the only difference is the cutting direction (i.e., [00$\overline{1}$] and [$\overline{1}$00]), there is a significant impact on the surface quality of the machined region due to the crystal anisotropy. We observed large brittle fractures in the [$\overline{1}$00] direction, whereas overall the [00$\overline{1}$] direction exhibited smooth surfaces. Figures~\ref{fig2}(f) and \ref{fig2}(g) show the variation in the critical depth of cut as a function of cutting direction on each plane. On the (001) plane, the critical depth variation was approximately 120~nm, and the lower bound value was 86~nm in the 270$^\circ$ direction ([0$\overline{1}$0] direction). On the other hand, the variation with the (010) plane was more significant than that with the (001) plane, and the lower bound also decreased (i.e., worsened). These considerable differences in critical depth of cut are consistent with surface observations, as shown in Fig.~\ref{fig2}(e).

We can understand the experimental results as follows.  The difference in critical depth of cut could be considered to originate from the slip system and the cleavage plane since they are strongly related to the ductile-brittle mode transition.  Cutting along the slip plane promotes ductile-mode cutting (i.e., plastic deformation), which contributes to the large critical depth of cut.   On the other hand, the cutting force against cleavage induces crystal parting where brittle fractures are easily manifested. They are explained intuitively in Fig.~\ref{fig2}(d).

\begin{figure*}[h]
	\centering
	\includegraphics[]{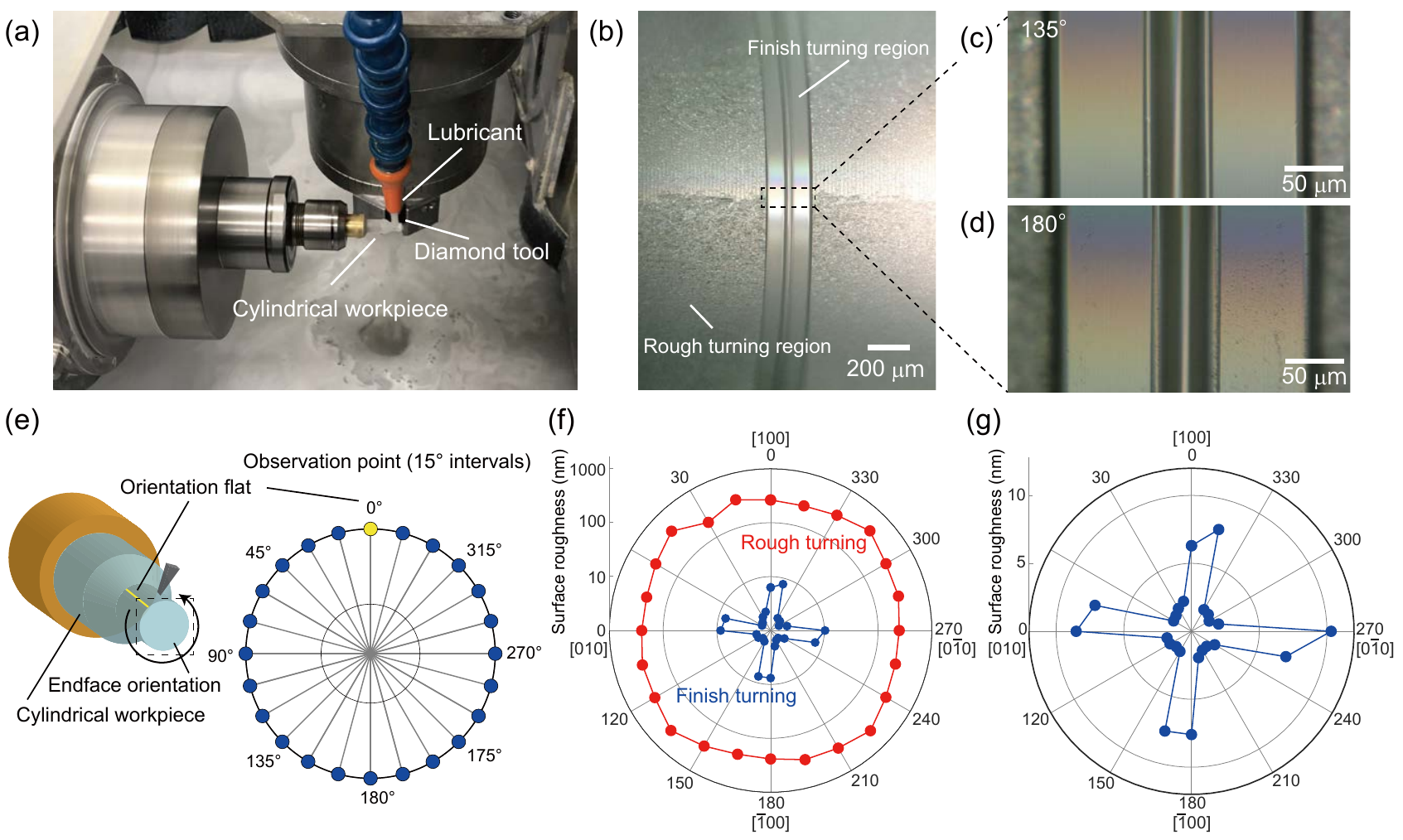}
	\caption{Ultra-precise cylindrical turning and surface roughness measurement of $\mathrm{MgF_2}$ single crystal. (a) Experimental setup of a ultra-precision lathe for the cylindrical turning of a single crystal. (b) Micrograph showing a machined surface, where clear boundaries are observed between the rough turning and finish turning regions. The horizontal boundary in the rough turning region is evidence of the dependence of the cutting direction on the crystal anisotropy in $\mathrm{MgF_2}$ single crystal. (c), (d) Magnified views of the finish turning region in 135$^\circ$ and 180$^\circ$, respectively. The machined surface of 135$^\circ$ is smoother than that of 180$^\circ$, which agrees with the result of the surface roughness (RMS) measurement. (e) Schematic of surface roughness measurement. The yellow line and dot correspond to orientation flat [100] with an endface orientation of (001). The surface roughness at a total of 24 points was measured at 15$^\circ$ intervals. (f) Measured surface roughness (RMS) of the finish turning (red dots) and rough turning (blue dots) regions. A quarter symmetry is clearly observed in the finish turning condition due to crystal anisotropy. (g) Magnified plot on linear scale of finish turning in (f).}
	\label{fig3}
\end{figure*}

The slip system and cleavage plane of single-crystal $\mathrm{MgF_2}$ are (110)[001] and (110), respectively; therefore the influence of cutting on the (001) plane on the critical depth of cut is less susceptible to the cutting direction because the cutting on the (001) plane is always normal to both the slip system and the cleavage plane.  With the (010) plane, however, we observed a large variation in the cutting direction, because the cutting periodically followed the same direction as the slip system (i.e., 0$^\circ$ and 180$^\circ$) as shown in Fig.~\ref{fig2}(g). In contrast, the smallest (i.e., worst) critical depth of cut was obtained for directions of 90$^\circ$ and 270$^\circ$.  They are in a configuration where the cutting force is applied in a direction almost perpendicular to the cleavage plane (110), as explained in Fig.~\ref{fig2}(d) (right panel). As a result, a shallow depth of cut is needed to obtain a ductile mode for these directions.  From the result of the orthogonal cutting experiment, we concluded that the depth of cut must be less than approximately 50~nm to maintain ductile mode cutting.

\section{Ultra-precision cylindrical turning}
\subsection{Procedure of cylindrical turning} 

Although the orthogonal cutting experiment provides information on the critical depth of cut for specific directions, the cutting direction continuously changes when cylindrical turning is performed to manufacture a crystalline cylinder workpiece. Therefore, the optimum turning parameters have to be investigated to achieve the smooth surface needed for a high-$Q$ microresonator. A $\mathrm{MgF_2}$ cylinder workpiece was prepared with an end-face orientation of (001) because a $z$-cut ($c$-cut) resonator is used to avoid optical birefringence. 

Cylindrical turning was performed using an ultra-precision aspheric surface machine (ULG-100E, TOSHIBA MACHINE), as shown in Fig.~\ref{fig3}(a). An $\mathrm{MgF_2}$ workpiece with a diameter of 6~mm was fixed to a brass jig, and then mounted on a vacuum chuck. The ultra-precision turning was conducted in the following three steps. Rough turning was initially undertaken to form the desired diameter (here 3~mm). It should be noted that this initial rough turning was performed in the brittle regime. Next, pre-finish cutting was conducted to remove the large cracks that occurred in brittle mode cutting with a removal thickness of 8~\textmu m.
Finally, finish cutting in the ductile mode completed the ultra-precise turning under the following cutting conditions: 500~min$^{-1}$ rotation speed, 0.1~mm/min feed rate, 50~nm depth of cut, and 2~\textmu m removal thickness (The cutting conditions in each step are detailed in Supplement~1). We can see that cracks deeper than 10~\textmu m that appeared during the rough turning could not be removed with pre-finish and finish cutting. Although the larger removal thickness results in lower production efficiency, it allows the removal of deep unwanted cracks.

The depth of cut at the finish turning step was set at 50~nm based on the result of the orthogonal cutting experiment. To achieve a smooth surface, other factors, such as the rotation speed, feed rate, and diamond tool, should be taken into account because these choices determine the effective cutting speed and cutting amount. In particular, previous studies have reported that the feed rate critically affects the quality of the machined surface as does the combination of the tool radius and depth of cut ~\cite{BLACKLEY199195,doi:10.1116/1.3071855,387fa24778d44dd2861a17056bc7dc42}. These studies draw attention to the fact that a fast feed rate induces brittle mode cutting if the depth of cut is kept below the critical value. Thus, we chose a slow feed rate ($\leq$~1~mm/min) when fabricating a smooth surface.

For the pre-finish and finish cutting, we used a single crystal diamond cutting tool with a 0.01~mm nose radius, a 0$^\circ$ rake angle, and a 10$^\circ$ clearance angle. In terms of the choice of the cutting tool, a smaller nose radius makes it possible to have a smaller contact area between tool and material during cylindrical turning, which helps to reduce any excess cutting force and leads to an improved surface quality. However, tools with a small nose radius are more fragile, which gives them a short lifespan; hence in this work we use different tools for the rough turning and finish turning stages.

The machined surfaces were observed using an optical microscope (VHX-5000, Keyence), as shown in Figs~\ref{fig3}(b)-\ref{fig3}(d). Clear boundaries can be identified in the micrograph images between the rough turning and finish turning regions.

\subsection{Surface roughness after cylindrical turning}

The surface roughness after cylindrical turning was measured using a scanning white-light interferometer (New View TM6200, Zygo) at 15$^\circ$ intervals from the orientation flat [100] defined as 0$^\circ$ [Fig.~\ref{fig3}(e)].  Figure~\ref{fig3}(f) and \ref{fig3}(g) show the cylindrical surface roughness with an end-face orientation of (001) of a $\mathrm{MgF_2}$ workpiece. Unsurprisingly, the surface roughness after the rough cutting exceeded 200~nm for the entire cylindrical surface, as shown with red dots in Fig.~\ref{fig3}(f).  The large roughness was caused by the brittle-regime cutting.  In contrast, the smoothness improved significantly after the finish cutting, which was performed under the ductile cutting condition [blue dots in Fig.~\ref{fig3}(f)].  The magnified plot on a linear scale is shown in Fig.~\ref{fig3}(g).  We confirmed that the turning condition for final cutting enabled us to achieve a smooth surface. Specifically, we obtained an RMS roughness of below 2~nm at 18 observation points.  The result also revealed an interesting feature of 90$^\circ$ periodicity, namely that specific observation points exhibited a slightly larger RMS roughness of 7.8~nm on average. Periodicity can also be seen in the micrograph shown in Fig.~\ref{fig3}(b)-\ref{fig3}(d); for instance, 135$^\circ$ exhibits a smoother machined surface than that in the 180$^\circ$ direction [Fig.~\ref{fig3}(c) and \ref{fig3}(d)].  This is evidence of the appearance of crystal anisotropy in $\mathrm{MgF_2}$ crystal, as observed in the orthogonal cutting experiment.

This periodicity can also be understood from the slip system and cleavage configurations shown in Fig.~\ref{fig2}(d).  The relatively rough surfaces can be explained in terms of specific directions where the excess cutting force acts on the boundaries of cleavage planes. The 15$^\circ$ asymmetry is due to the rotation direction of the workpiece; the force on the cleavage plane exerts stress only in the clockwise direction (the opposite direction to workpiece rotation). There is nevertheless excellent surface integrity as long as the cutting remains on a cylindrical surface where the cutting force circumvents the crystal anisotropy. (A detailed discussion of the effect of crystal structure on cutting condition is presented in Supplement~1.)

The cylindrical turning described here is a pre-process in microresonator fabrication. It should be noted that the measured roughness is the result of the implosion of a machined cylinder, not the dimensional resonator surface. However, the results we obtained allow us to predict the surface of the resonator under the employed cutting conditions.

\subsection{Fabrication and cleaning of microresonators}

\begin{figure}[h]
	\centering
	\includegraphics[]{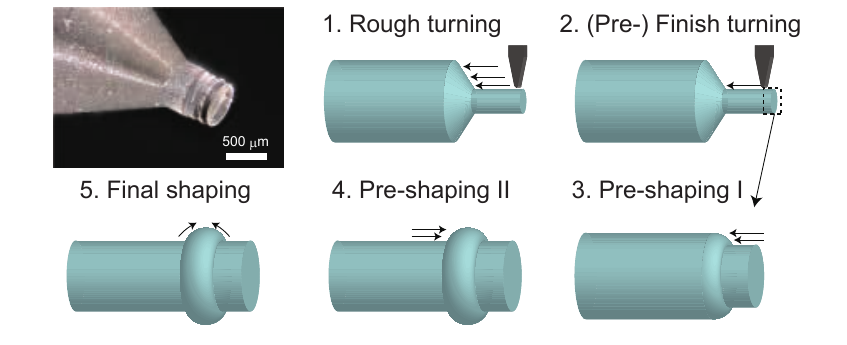}
	\caption{Fabrication flow of WGM microresonator when using ultra-precision turning. First, a rough turning determines the approximate diameter of resonator. Next, pre-finish and finish turning with ductile mode cutting are used to realize a cylindrical surface that is smooth and entirely crack-free. Finally, fully-programmed shaping steps are performed to fabricate the designed resonator structure.}
	\label{fig4}
\end{figure}

Crystalline microresonators are fabricated using the same ultra-precision machine as that used in the cylindrical turning experiment (ULG-100E, TOSHIBA MACHINE). The resonator diameter is determined after the finish cutting. It should be noted that the diameter can be precisely controlled by measuring the diameter and undertaking additional turning prior to microresonator fabrication. The resonator shape is carefully fabricated by feeding a diamond tool under the critical cutting depth. Here, the turning motion is fully and automatically controlled by the NC program. The manufacturing procedure is shown in Fig.~\ref{fig4}, and the total fabrication time is about ten hours. We determined the cutting condition based on the cylindrical turning experiment and employed a finish turning condition at the resonator shaping step (Also see Supplement~1). 

Once we had completed the fabrication, we cleaned the microresonator to remove lubricant and small chips attached to the surface. We emphasize that proper cleaning is essential for obtaining a high-$Q$ as well as optimized cutting conditions. Since we used water-soluble oil as a machining lubricant during the ultra-precision turning, we first used acetone solution to clean the microresonator surface and remove the remaining lubricant. A lens cleaning tissue is usually used to wipe the resonator, but there is the possibility that it might scratch or damage the resonator surface, which could be a critical problem in terms of degrading the $Q$-factor. Alternatively, to avoid unwanted damage on the resonator, we can employ ultrasonic cleaning. (The cleaning method is detailed in Supplement~1.) The use of an ultrasonic cleaner enables us to clean the surface without touching or rubbing it.  It is also a great advantage for fully automated fabrication combined with ultra-precision turning.

\section{Q-Factor and dispersion of fabricated microresonators}

\begin{figure*}[t]
	\centering
	\includegraphics[]{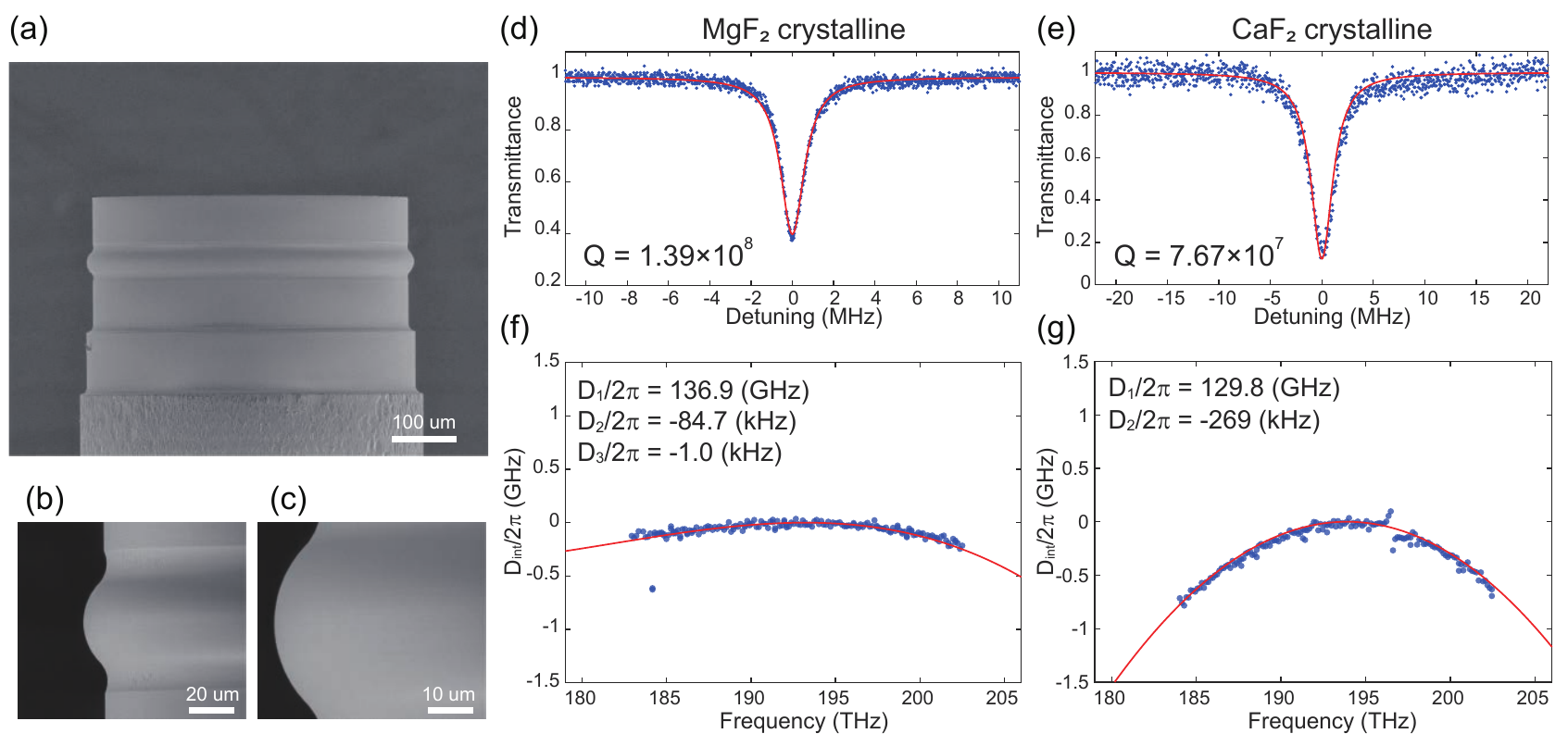}
	\caption{$Q$-factor and dispersion measurement of crystalline microresonators fabricated by ultra-precise turning. (a) SEM image of a fabricated $\mathrm{MgF_2}$ microresonator with a diameter of 508~\textmu m and a curvature radius of 36~\textmu m. (b), (c) Magnified views of the resonator. (d) Normalized transmission spectra of the fabricated $\mathrm{MgF_2}$ microresonator. The Lorentzian fitting (red line) yield loaded a $Q$ value of 139 million. (e) Normalized transmission spectra of the fabricated $\mathrm{CaF_2}$ microresonator. The fitting curves give a loaded $Q$ for the fundamental mode of 76.7 million. (f), (g) Measured dispersion $D_\mathrm{int}$ versus frequency. The red curve indicates the calculated dispersion of the fundamental TM mode, which agrees well with the experimental result.}
	\label{fig5}
\end{figure*}

The $Q$-factor and dispersion were measured in crystalline microresonators fabricated with the procedure described above. We fabricated an $\mathrm{MgF_2}$ WGM resonator and a $\mathrm{CaF_2}$ WGM resonator with the same curvature radii of 36~\textmu m. The diameters were 508~\textmu m for the $\mathrm{MgF_2}$ resonator and 512~\textmu m for $\mathrm{CaF_2}$ resonator. Figures~\ref{fig5}(a)-\ref{fig5}(c) show SEM images of the fabricated $\mathrm{MgF_2}$ microresonator. Although the two resonators were fabricated with the same motion program and cutting conditions, their diameters differ slightly as a result of differences in the positioning accuracy in the cylindrical turning process and the original diameter of the cylindrical workpiece. As described in the previous section, the additional measurement of the workpiece dimension enables us to achieve the practical precise control of the diameter at the sub-micrometer level.

Figures~\ref{fig5}(d) and \ref{fig5}(e) show the measured transmission spectra of the fabricated microresonators. We launched light from a frequency tunable laser source, which was coupled into the resonator via a tapered optical silica fiber. A polarization controller was used to adjust the polarization before the light coupling. The transmitted light was monitored with a high-speed photodetector and oscilloscope, where we used a calibrated fiber Mach-Zehnder interferometer as the frequency reference. 
The full-width at half-maximum (FWHM) linewidth of the $\mathrm{MgF_2}$ resonator was 1.40~MHz, which corresponds to a loaded $Q=1.39\times10^8$ at a wavelength of 1545~nm. Also, the $\mathrm{CaF_2}$ resonator had a linewidth of 2.53~MHz at 1546~nm, corresponding to $Q=7.67\times10^7$. We measured the $Q$-factor in different wavelengths regions and recorded comparably high-$Q$ values for other resonant modes. The obtained $Q$, which exceeded 100~million, is the highest value recorded in a crystalline WGM microresonator fabricated solely by ultra-precision machining without a conventional polishing process. In other words, our approach has overcome the manufacturing limitation, namely the need for skilled manual techniques throughout the fabrication process to obtain ultrahigh-$Q$ crystalline microresonators.

Figures~\ref{fig5}(f) and \ref{fig5}(g) show the measured integrated dispersion, defined as $D_{\mathrm{int}}=\omega_\mu-\omega_0-D_1\mu=D_2\mu^2/2+D_3\mu^3/6+\cdots$, where $\omega_\mu/2\pi$ is the resonance frequency of the $\mu$-th mode ($\mu=0$ designates the center mode), $D_1/2\pi$ is the equidistant free-spectral range (FSR), $D_2/2\pi$ is the second-order dispersion linked to group velocity dispersion, and the above $D_3/2\pi$ terms correspond to higher-order dispersion. The microresonator dispersion measurement was performed assisted by a fiber laser comb and a wavelength meter~\cite{FujiiNano}. The measured dispersion agrees well with the theoretical dispersion calculated with the finite element method (FEM) by using (COMSOL Multiphysics), and these results indicate that ultra-precision turning enables us to obtain the designed resonator shape.   Measured results and the fabrication flow of a WGM crystalline resonator with a sophisticated cross-sectional shape, i.e., a triangular shape, are provided in Supplement~1.

In fact, we realized a Kerr frequency comb in the fabricated $\mathrm{MgF_2}$ crystalline microresonator by machining alone where the dispersion was designed to generate octave-wide parametric oscillation~\cite{Fujii:19}. Hence, fully computer-controlled machining could be a great advantage as regards extending the potential of crystalline microresonators for optical frequency comb generation from the standpoint of dispersion engineering.

\section{Discussion}

\subsection{Outlook: Towards further Q-factor improvement}

The total (loaded) $Q$-factor is determined by folding the several loss contributions as,
\begin{equation}
	Q^{-1}_{\mathrm{tot}} = Q^{-1}_{\mathrm{mat}} + Q^{-1}_{\mathrm{surf}} + Q^{-1}_{\mathrm{scatt}} + Q^{-1}_{\mathrm{rad}} + Q^{-1}_{\mathrm{ext}},
\end{equation}
where $Q^{-1}_{\mathrm{mat}}$ is determined by material absorption,  $Q^{-1}_{\mathrm{surf}}$ and $Q^{-1}_{\mathrm{scatt}}$ are determined by surface absorption and scattering loss, respectively. The radiation (tunneling) loss is given by $Q^{-1}_{\mathrm{rad}}$, and  $Q^{-1}_{\mathrm{ext}}$ is related to the coupling rate between the resonator and the external waveguide (e.g., tapered fiber, prism). 
Since the total $Q$-factor can readily reach $10^9$ by polishing in fluoride crystalline resonators, the effect of $Q^{-1}_{\mathrm{mat}}$, $Q^{-1}_{\mathrm{rad}}$, and $Q^{-1}_{\mathrm{ext}}$ should be negligible. $Q^{-1}_{\mathrm{surf}}$ is one possible reason for this, whereas single crystals such as $\mathrm{MgF_2}$ and $\mathrm{CaF_2}$ inhibit the diffusion of water into the crystal lattice, which makes $Q^{-1}_{\mathrm{surf}}$ negligible in our case~\cite{GRUDININ200633}.

\begin{figure}[h]
	\centering
	\includegraphics[]{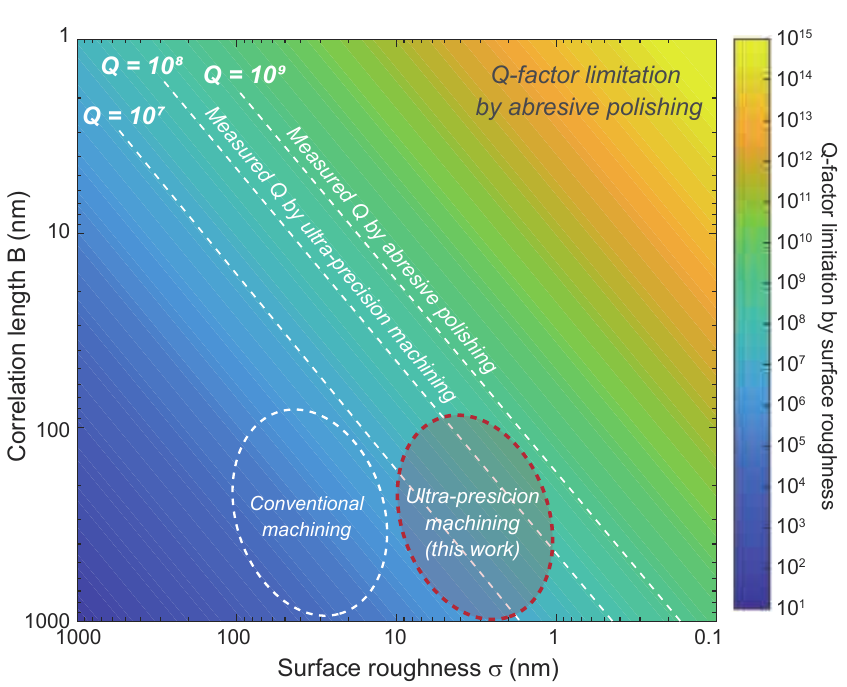}
	\caption{$Q$-factor limitation caused by surface scattering loss in $\mathrm{MgF_2}$ crystalline microresonators derived from Eq.~(\ref{Eq:Q_scatt}) ($\lambda$=1550~nm and $R=250$~\textmu m). The dashed contours show the estimated value from our measurement and previous studies~\cite{Gorodetsky:00,GRUDININ200633}.}
	\label{fig6}
\end{figure}

Then, we highlight $Q^{-1}_{\mathrm{scatt}}$ as a fundamental limitation of ultra-precision machining. Since the surface roughness of the polished resonator reaches a sub-nanometer scale~\cite{GRUDININ200633}, it is reasonable to consider that the surface scattering limits $Q$ in diamond turning (Fig.~\ref{fig3}(g) shows a surface roughness of a few nanometer scale). The maximum $Q$-factor as regards surface roughness can be estimated as~\cite{Gorodetsky:00,GRUDININ200633}:

\begin{equation}
	Q_{\mathrm{scatt}} \approx \frac{3 \lambda^3 R}{8 n \pi^2 B^2 \sigma^2} \label{Eq:Q_scatt}
\end{equation}
where $R$ is the resonator radius, $n$ is the refractive index, $B$ is the correlation length, and $\sigma$ is the surface roughness (RMS). The maximum $Q$-factor versus surface roughness and correlation length of $\mathrm{MgF_2}$ resonator is plotted in Fig.~\ref{fig6}. Theoretically achievable values for ultra-precision machining correspond to $Q$ values of $10^7-10^9$, which are consistent with measured $Q$-factors. The plot indicates that the roughness of the machined surface could limit the present $Q$. A possible way to improve the surface roughness and correlation length is to optimize the cutting parameters, for example by using a smaller depth of cut and a lower feed rate. Specifically, ideal conditions are believed to realize an ultrasmooth surface for the entire cylindrical position, and consequently eliminate the effect of crystal anisotropy, as seen in Fig.~{\ref{fig3}}(g). 

In addition, we should take the effect of subsurface damage into account since it causes the degeneration of the inner structure of the material. Subsurface damage occurs when machining a single crystal, and so it has been intensively studied in the field of micromachining and material science~\cite{YAN2009378}. Such underlayer damage could degrade $Q$ in the same way as surface scattering; therefore, we investigated the surface and subsurface damage by using a SEM and a transmission electron microscope (TEM) in comparison with the results for polishing. As a result, we found that the damaged subsurface layers were around several tens of nanometers with single-crystal precise turning. (Details and results are presented in Supplement~1.)  It is generally known that the subsurface damage mechanism strongly depends on the crystal properties and cutting condition, and the efforts to reduce the subsurface damage are described elsewhere~\cite{MIZUMOTO201873}. The reduction of underlayer damage could also help to improve the present $Q$.

\section{Conclusion}

In conclusion, we demonstrated the fabrication of an ultrahigh-$Q$ crystalline microresonator by using ultra-precision turning alone. For the first time, we achieved a $Q$ value exceeding 100 million without polishing and thereby managed both an ultrahigh-$Q$ and dispersion engineering simultaneously. We revealed the critical depth of cut needed to sustain ductile mode cutting, and this information contributes significantly to reducing surface roughness. Moreover, we proposed an optimal cutting condition for cylindrical turning for realizing an ultrasmooth surface throughout an entire cylindrical surface. This result provides the path towards the fabrication of a high-$Q$ microresonator without the need for skilled manual work. Furthermore, we discussed the possibility of further improving the $Q$-factor from the standpoint of the theoretical limitation imposed by surface roughness. The described fabrication and cleaning procedure can be applied to various single-crystal materials and will raise the potential for realizing crystalline microresonators.

\section*{Funding Information}
Japan Society for the Promotion of Science (JSPS) (JP18J21797) Grant-in-Aid for JSPS Fellow; JSPS KAKENHI (JP18K19036); Strategic Information and Communications RD Promotion Programme (191603001) from MIC.

\section*{Acknowledgments}

The authors thank Dr.~Y.~Mizumoto for fruitful discussion, and also H.~Amano, T.~Takahashi, M.~Fuchida, and K.~Wada for technical support.
\\

See Supplement 1 for supporting content.

\section*{Disclosures}

The authors declare no conflicts of interest.


%


\end{document}